\documentclass[runningheads]{llncs}

\usepackage{graphicx}
\usepackage[algoruled, portuguese]{algorithm2e}  
\usepackage{amsmath}
\usepackage{tikz}

\usepackage{caption}
\usepackage{subcaption}

\usepackage[hidelinks]{hyperref}

\usepackage{xcolor}
\usepackage{soulutf8}

\usepackage[portuges]{babel}

\begin{document}
\title{Uma extensão de Raft com propagação epidémica\thanks{Este trabalho é cofinanciado pela Componente 5 - Capitalização e Inovação Empresarial, integrada na Dimensão Resiliência do Plano de Recuperação e Resiliência no âmbito do Mecanismo de Recuperação e Resiliência (MRR) da União Europeia (EU), enquadrado no Next Generation UE, para o período de 2021 - 2026, no âmbito do projeto ATE, com a referência 56.}$^,$\thanks{Publicado nas atas do INForum'2023: \url{https://dei.fe.up.pt/inforum23}.}}
\author{André Gonçalves\orcidID{0009-0002-1284-094X} \and
Ana Nunes Alonso\orcidID{0000-0002-0519-9675} \and
José Pereira\orcidID{0000-0002-3341-9217} \and
Rui Oliveira\orcidID{0000-0003-3408-7346}}
\institute{INESC TEC e Universidade do Minho, Braga, Portugal}
\maketitle

\begin{abstract}
O algoritmo de acordo Raft é reconhecido pela sua facilidade de compreensão e implementação prática, sendo atualmente adotado em sistemas como o Kubernetes. No entanto, tem algumas limitações em termos de escalabilidade e 
desempenho
por concentrar o esforço no líder.
Neste trabalho apresentamos um novo algoritmo que expande o Raft com a incorporação de mecanismos de propagação epidémica para descentralizar o esforço da replicação.
A nossa proposta é avaliada experimentalmente com uma implementação em Go e testada com um número significativo de processos.

\keywords{Acordo distribuído \and Raft \and Propagação epidémica.}
\end{abstract}

\section{Introdução}
O acordo é um dos problemas mais fundamentais em sistemas distribuídos tolerantes a faltas. 
O problema surge da necessidade de processos distribuídos chegarem a um acordo sobre um certo valor proposto, que se torna especialmente complexo na presença de faltas. 
Bastantes algoritmos foram desenvolvidos, ao longo das últimas décadas, para resolver este problema, cada um com os seus pressupostos e garantias. 
Entre eles destaca-se o Paxos \cite{paxos}, desenvolvido por Lamport, sob a forma de uma máquina de estados replicada \cite{smr}, procurando chegar a acordo sobre a ordem em que comandos serão aplicados. 
No entanto, Paxos é considerado difícil de compreender, o que motivou o desenvolvimento do algoritmo de acordo Raft \cite{raft}.

Raft garante que um sistema de máquinas de estados funcione como uma única entidade, desde que a maioria dos processos esteja operacional e que mensagens entre estes sejam inevitavelmente entregues. 
O algoritmo alcança a sua simplicidade ao decompor-se em subproblemas e ao usar liderança forte, onde um processo eleito (líder) assume a responsabilidade de tomar as decisões para todo o sistema. 
Em operação normal, o líder recebe os pedidos dos clientes, adiciona-os ordenadamente a um registo, que replica pelos outros processos (seguidores). Assim que uma maioria de seguidores informa o líder de que o registo foi replicado com sucesso até um certo ponto, este pode aplicar as entradas até esse ponto e responder aos clientes respetivos. 
A centralização do algoritmo no líder torna-o simples, mas, em contrapartida, pouco escalável: com o aumento do número de réplicas ou carga de trabalho pelos clientes, o líder rapidamente esgota os seus recursos, e torna-se o gargalo do sistema.

Propomos a adição de novos mecanismos baseados em propagação epidémica \cite{gossip} ao Raft para superar estas limitações. Com propagação epidémica os processos comunicam através do encaminhamento de mensagens entre os seus vizinhos.
A nossa abordagem consiste em introduzir no líder rondas de propagação epidémica periódicas, enviadas numa permutação \cite{mutable}, para replicar o registo.
Com esta abordagem, as mensagens do líder conseguem alcançar todos os seguidores, mesmo que o líder não esteja diretamente conectado com cada um deles. Desta forma, evitamos que eleições desnecessárias aconteçam, que em Raft são iniciadas por um seguidor quando este não recebe comunicação do líder durante um determinado limite de tempo.
Adicionalmente, estudamos a utilização de novas estruturas de dados na propagação epidémica, que permitirão a qualquer processo descobrir o próximo índice do registo até onde as entradas estão confirmadas e poderão ser aplicadas. 

Implementamos o algoritmo Raft com e sem os novos mecanismos em Paxi, uma infraestrutura de código aberto para prototipagem e avaliação de algoritmos de replicação escrita em Go\,\cite{paxi,paxiarticle}. As implementações são avaliadas num sistema com 51 processos, cada com um CPU dedicado.

\section{Contexto}

\begin{figure}[t]
    \centering
    \tikzset{every picture/.style={line width=0.75pt}} 

\scalebox{.9}{
\begin{tikzpicture}[x=0.75pt,y=0.75pt,yscale=-1,xscale=1]

\draw  [fill={rgb, 255:red, 206; green, 247; blue, 164 }  ,fill opacity=1 ] (261.16,206) .. controls (261.16,188.46) and (275.38,174.24) .. (292.91,174.24) .. controls (310.45,174.24) and (324.67,188.46) .. (324.67,206) .. controls (324.67,223.53) and (310.45,237.75) .. (292.91,237.75) .. controls (275.38,237.75) and (261.16,223.53) .. (261.16,206) -- cycle ;
\draw  [fill={rgb, 255:red, 255; green, 170; blue, 180 }  ,fill opacity=1 ] (425.16,206) .. controls (425.16,188.46) and (439.38,174.24) .. (456.91,174.24) .. controls (474.45,174.24) and (488.67,188.46) .. (488.67,206) .. controls (488.67,223.53) and (474.45,237.75) .. (456.91,237.75) .. controls (439.38,237.75) and (425.16,223.53) .. (425.16,206) -- cycle ;
\draw  [fill={rgb, 255:red, 168; green, 243; blue, 223 }  ,fill opacity=1 ] (85,206) .. controls (85,188.51) and (99.18,174.33) .. (116.67,174.33) .. controls (134.16,174.33) and (148.33,188.51) .. (148.33,206) .. controls (148.33,223.49) and (134.16,237.67) .. (116.67,237.67) .. controls (99.18,237.67) and (85,223.49) .. (85,206) -- cycle ;
\draw    (64.55,158.35) .. controls (70.57,174.59) and (74.21,177.14) .. (85.59,183.43) ;
\draw [shift={(87.25,184.35)}, rotate = 208.86] [color={rgb, 255:red, 0; green, 0; blue, 0 }  ][line width=0.75]    (10.93,-3.29) .. controls (6.95,-1.4) and (3.31,-0.3) .. (0,0) .. controls (3.31,0.3) and (6.95,1.4) .. (10.93,3.29)   ;
\draw    (258.48,207.58) -- (153.98,207.58) ;
\draw [shift={(151.98,207.58)}, rotate = 360] [color={rgb, 255:red, 0; green, 0; blue, 0 }  ][line width=0.75]    (10.93,-3.29) .. controls (6.95,-1.4) and (3.31,-0.3) .. (0,0) .. controls (3.31,0.3) and (6.95,1.4) .. (10.93,3.29)   ;
\draw    (131.48,173.58) .. controls (153.65,149.45) and (251,146.66) .. (279.24,172.39) ;
\draw [shift={(280.48,173.58)}, rotate = 225.54] [color={rgb, 255:red, 0; green, 0; blue, 0 }  ][line width=0.75]    (10.93,-3.29) .. controls (6.95,-1.4) and (3.31,-0.3) .. (0,0) .. controls (3.31,0.3) and (6.95,1.4) .. (10.93,3.29)   ;
\draw    (305.98,173.08) .. controls (328.64,147.97) and (413.63,148.82) .. (443.78,172.98) ;
\draw [shift={(445.12,174.1)}, rotate = 221.15] [color={rgb, 255:red, 0; green, 0; blue, 0 }  ][line width=0.75]    (10.93,-3.29) .. controls (6.95,-1.4) and (3.31,-0.3) .. (0,0) .. controls (3.31,0.3) and (6.95,1.4) .. (10.93,3.29)   ;
\draw    (432.12,233.1) .. controls (354.51,263.95) and (227.4,268.06) .. (137.47,236.58) ;
\draw [shift={(136.12,236.1)}, rotate = 19.57] [color={rgb, 255:red, 0; green, 0; blue, 0 }  ][line width=0.75]    (10.93,-3.29) .. controls (6.95,-1.4) and (3.31,-0.3) .. (0,0) .. controls (3.31,0.3) and (6.95,1.4) .. (10.93,3.29)   ;
\draw    (324.12,192.1) .. controls (358.59,174.37) and (358.14,240.08) .. (324.67,221.99) ;
\draw [shift={(323.12,221.1)}, rotate = 30.96] [color={rgb, 255:red, 0; green, 0; blue, 0 }  ][line width=0.75]    (10.93,-3.29) .. controls (6.95,-1.4) and (3.31,-0.3) .. (0,0) .. controls (3.31,0.3) and (6.95,1.4) .. (10.93,3.29)   ;

\draw (262.67,198.67) node [anchor=north west][inner sep=0.75pt]  [font=\footnotesize] [align=left] {Candidato};
\draw (441.67,199.33) node [anchor=north west][inner sep=0.75pt]  [font=\footnotesize] [align=left] {Líder};
\draw (90.67,199) node [anchor=north west][inner sep=0.75pt]  [font=\footnotesize] [align=left] {Seguidor};
\draw (69.88,143.57) node  [font=\scriptsize] [align=left] {(1) inicia ou \\recupera};
\draw (155.45,179.35) node [anchor=north west][inner sep=0.75pt]  [font=\scriptsize] [align=left] {(4) perde eleição ou\\descobre termo maior };
\draw (148.45,141.35) node [anchor=north west][inner sep=0.75pt]  [font=\scriptsize] [align=left] {(2) tempo limite passa};
\draw (296.45,141.35) node [anchor=north west][inner sep=0.75pt]  [font=\scriptsize] [align=left] {(3) recebe uma maioria de votos};
\draw (225.45,261.35) node [anchor=north west][inner sep=0.75pt]  [font=\scriptsize] [align=left] {(6) descobre termo maior};
\draw (351.45,193.35) node [anchor=north west][inner sep=0.75pt]  [font=\scriptsize] [align=left] {(5) tempo \\limite passa};

\end{tikzpicture}
}
    \caption{Transições entre estados}
    \label{fig:estados}
\end{figure}
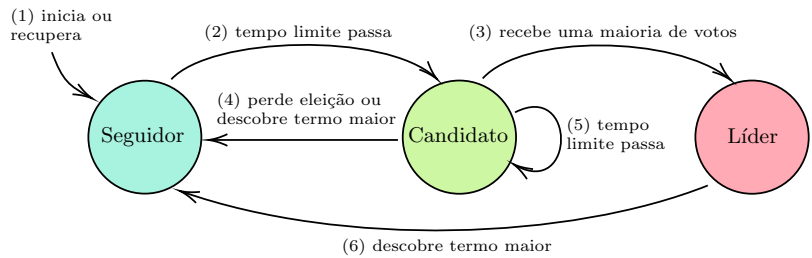

Raft garante a coerência de máquinas de estados através da gestão de um registo de operações replicado.
Num sistema Raft, os processos assumem, num determinado momento, um de três estados, conforme a Fig.~\ref{fig:estados}.
Um processo inicia-se no estado de \emph{seguidor}, que recebe pedidos e responde de acordo, nunca iniciando a comunicação. Este transita para o estado de \emph{candidato} quando assume que o líder falhou, iniciando uma eleição e solicitando ativamente votos aos outros processos para ser eleito. Ao receber uma maioria de votos transita para o estado de \emph{líder}, assumindo toda a responsabilidade de manter o registo replicado coerente e processar os pedidos dos clientes. Referimos-nos a cada processo pelo estado em que se encontra. 

No contexto do Raft, o tempo é dividido logicamente em \emph{termos}. Ao termo é associado um número natural que cresce monotonamente, que agindo como um relógio lógico, permite a ordenação total de eventos. O termo dum processo é incrementado quando este inicia (ou reinicia) uma eleição, ou quando aprende de outro processo que o seu termo está desatualizado, ou seja, é menor.

A interação entre processos é efetuada por invocações remotas (\textit{Remote Procedure Calls} / RPCs) assíncronas. Ao iniciar uma invocação remota o processo não bloqueia, permitindo-lhe processar outros pedidos que cheguem entretanto. As invocações são reexecutadas após um tempo limite, para tolerar a perda de mensagens na rede. São necessários dois tipos de RPC: \texttt{RequestVote RPC}, iniciados por candidatos para recolher votos numa eleição; e \texttt{AppendEntries RPC}, iniciados pelo líder para replicar o seu registo, servindo também como \textit{heartbeat}, para informar os seguidores de que está ativo.

Raft decompõe a sua lógica em três subproblemas: eleição do líder, replicação do registo e garantia de acordo. Uma vez que a nossa proposta se foca apenas na replicação, detalhamos aqui apenas a replicação de registo.
Após ser eleito líder, este deve: processar pedidos dos clientes, replicar o seu registo e enviar \textit{hearbeats} para manter a liderança. Os clientes enviam operações para ser aplicadas na máquina de estados replicada. Ao receber um pedido de um cliente, o líder adiciona a operação e termo de receção como uma nova entrada no seu registo. Depois, emite RPCs \texttt{AppendEntries} para replicar a nova entrada para cada seguidor. Assim que o líder sabe que a entrada foi replicada com sucesso para uma maioria de processos, ou seja, foi \emph{confirmada}, pode aplicar a respetiva operação na sua máquina de estados e responder com o resultado ao respetivo cliente. 

Cada processo tem um valor que cresce monotonamente, \texttt{CommitIndex}, que corresponde ao índice no registo da última entrada confirmada. Todas as entradas anteriores ao \texttt{CommitIndex} são também consideradas confirmadas. O líder informa via \texttt{AppendEntries} do seu \texttt{CommitIndex} para que os seguidores possam atualizar os seus próprios. Os processos devem aplicar ordenadamente as novas entradas confirmadas na sua máquina de estados.

O termo de receção e o índice em cada entrada são necessários para detetar incoerência entre registos.
O líder incluí nos pedidos \texttt{AppendEntries} uma partição do seu registo com as entradas ainda não replicadas com sucesso no respetivo processo destinatário.
Ao receber um pedido \texttt{AppendEntries}, o seguidor verifica se a entrada anterior às entradas no pedido não entra em conflito com o seu registo e, nesse caso, pode adicionar as novas entradas. Na resposta, o seguidor informa se as entradas foram replicadas com sucesso.
No Raft a propriedade de correspondência de registo permite reconhecer se dois registos são iguais até um dado índice. A propriedade diz o seguinte: se dois registos contêm no mesmo índice uma entrada com o mesmo termo, então todos as entradas até esse índice são idênticas.
Ao receber a indicação de sucesso na resposta, o líder reconhece que o registo foi replicado com sucesso até ao maior índice das entradas enviadas no respetivo pedido RPC. Caso receba indicação de insucesso, significa que não enviou entradas suficientes no pedido e deve reenviar um novo RPC \texttt{AppendEntries} com entradas começando num ponto anterior. O líder poderá receber indicação de insucesso várias vezes até chegar a um ponto onde os registos sejam compatíveis.

\section{Extensão com Propagação Epidémica}

Raft assume conectividade de todos-para-todos entre os processos do sistema desde que a rede não esteja particionada. O líder replica o seu registo via mensagens um-para-um para os seguidores. A nossa ideia consiste em disseminar mensagens para replicação (pedidos \texttt{AppendEntries}) no sistema com propagação epidémica, que fornece escalabilidade e robustez mesmo em redes em que a conectividade não é transitiva. Além disso, expandimos ainda mais o algoritmo ao introduzir estruturas de dados no estado dos processos, partilhadas com a propagação epidémica, que permitirá avançar o \texttt{CommitIndex} de forma descentralizada. 

\subsection{Propagação Epidémica de \texttt{AppendEntries}}
Raft usa RPCs para comunicação entre processos, mas enquanto esta abordagem torna o algoritmo mais simples também o limita. Na nova versão de Raft, o líder dissemina periodicamente o mesmo pedido \texttt{AppendEntries} por propagação epidémica e quando um seguidor recebe um pedido pela primeira vez, responde ao respetivo líder. Contudo, pode acontecer que processos não recebam um pedido numa dada ronda de propagação e, portanto, falhar a replicação de entradas. Então, no próximo pedido de \texttt{AppendEntries} recebido, este irá responder que a replicação falhou e o líder iniciará a invocação de RPCs \texttt{AppendEntries} com o processo até que as entradas em falta sejam replicadas com sucesso.

\begin{algorithm}[b]
\SetAlgoLined
\KwData{\\\DataSty{F} $\leftarrow$ fanout value\\
    \DataSty{c} $\leftarrow$ 0\\
    \DataSty{u} $\leftarrow$ permutação de $1$ a $n$ exceto $i$}

\SetKwFunction{Ronda}{Ronda}
\SetKwProg{Fn}{Function}{:}{}
\Fn{\Ronda{\ArgSty{m}}}{
    \For{$i \leftarrow 0$ \KwTo $F$}{
        \textbf{enviar} \ArgSty{m} para \DataSty{u}[(\DataSty{c} + $i$)$\mod$ \DataSty{F}]\;
    } 
    \DataSty{c} $\leftarrow$ \DataSty{c} + \DataSty{F}\;
}
\caption{Ronda de propagação epidémica para o processo $P_i, i \in 0..n-1$}
\label{alg:permutgossip}
\end{algorithm}

O líder transmite pedidos \texttt{AppendEntries} em rondas de propagação epidémica numa permutação \cite{mutable} para replicar as entradas que ainda não foram confirmadas. O líder calcula uma permutação de seguidores, isto é uma lista aleatória dos identificadores dos seguidores, que percorre circularmente em cada ronda de propagação epidémica, como especificado no Algoritmo~\ref{alg:permutgossip}. As rondas são iniciadas periodicamente para replicação e para servir de \textit{heartbeat}. 
O líder inicia periodicamente uma ronda de propagação epidémica de um pedido \texttt{AppendEntries} com as entradas ainda não confirmadas. Por outro lado, se todas as entradas do registo foram confirmadas, o líder pode escolher um intervalo de tempo maior para iniciar rondas com \textit{heartbeat}, para manter a liderança.

\begin{figure}[t]
    \centering
    \makebox[\textwidth]{
\includegraphics[width=1\textwidth]{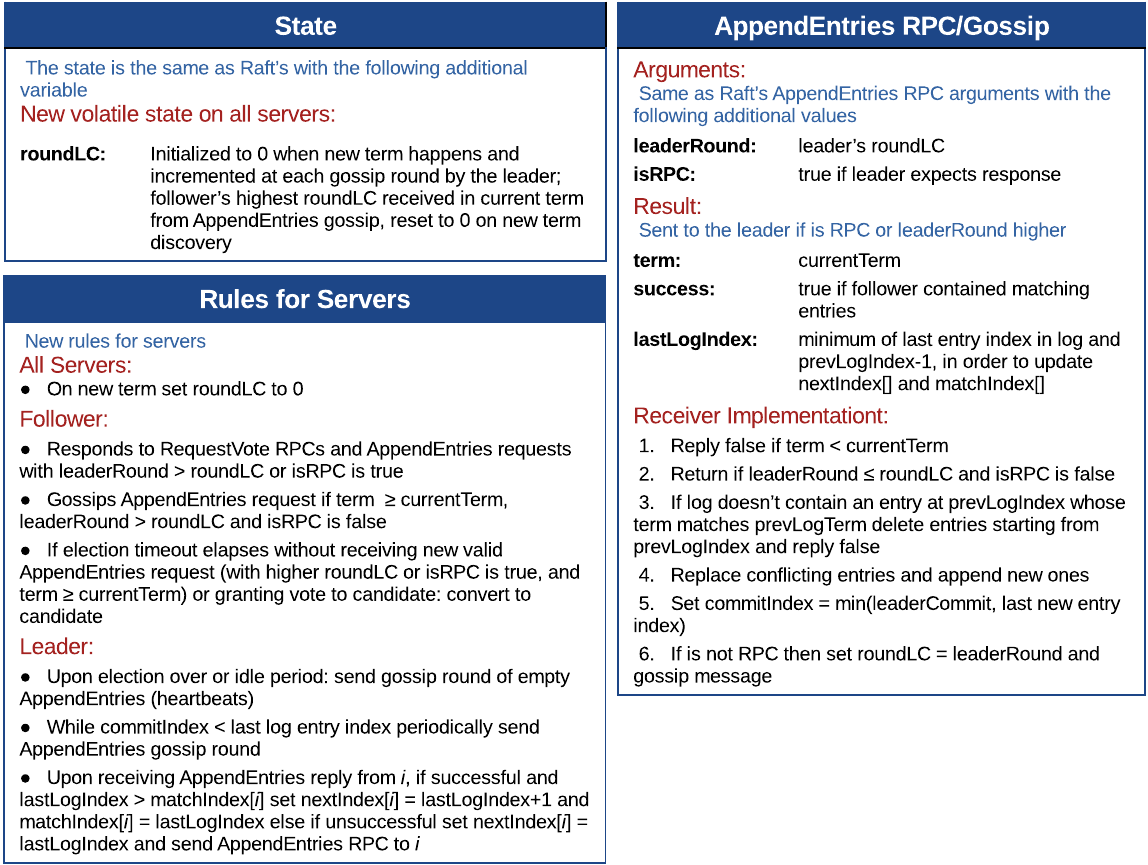}
    }
    \caption{Raft com propagação epidémica de \texttt{AppendEntries}.}
    \label{fig:driftwood}
\end{figure}

Em pedidos \texttt{AppendEntries} o líder incluí uma variável booleana que permite aos seguidores distinguir entre pedidos \texttt{AppendEntries} provenientes de RPC ou de propagação epidémica. Caso seja um RPC emitido pelo líder, o seguidor tem que responder ao pedido. Se o pedido vem de propagação epidémica então o seguidor responde apenas se for a primeira vez que o recebe.

Para distinguir se os pedidos \texttt{AppendEntries} são provenientes de uma ronda de propagação epidémica mais recente, os processos mantêm no seu estado uma variável inteira, \texttt{RoundLC}, que serve como um relógio lógico no mandato atual. Cada processo repõe o seu \texttt{RoundLC} a zero quando o mandato muda. O líder incrementa o seu \texttt{RoundLC} quando começa uma nova ronda e inclui esta variável nos pedidos \texttt{AppendEntries} para que os seguidores a possam atualizar no seu estado. Assim, seguidores evitam processar e propagar mensagens antigas ou já recebidas. Seguidores consideram as mensagens mais recentes (com \texttt{RoundLC} maior que o seu) um \textit{heartbeat} do líder.

O nosso algoritmo preserva grande parte da lógica do Raft original. A Fig.~\ref{fig:driftwood} sumariza os novos mecanismos adicionados e as alterações necessárias para expandir o Raft, mantendo a língua original para facilitar a comparação.

\subsection{Novas Estruturas de Dados}

A nova versão de Raft diminui o número de mensagens que o líder tem de trocar ao enviar entradas não confirmadas em lote e ao deixar a disseminação destas entradas para os seguidores. Contudo, mantemos a abordagem de que o líder só avança o \texttt{CommitIndex} após receber confirmação de replicação de uma maioria de seguidores.
Para avançar o valor do \texttt{CommitIndex} de forma descentralizada acrescentamos ao Raft novas estruturas de dados para a propagação epidémica. Estas estruturas consistem em três variáveis utilizadas da seguinte forma:
\begin{itemize}
    \item \texttt{\textbf{Bitmap:}} mapa de bits, no qual apenas o processo pode alterar o respetivo bit para ``um"; regista a votação para \texttt{MaxCommit}
    \item \texttt{\textbf{MaxCommit:}} valor máximo que pode ser atribuído a \texttt{CommitIndex}, corresponde ao maior índice confirmado observado pelo processo; 
    \item \texttt{\textbf{NextCommit:}} valor a ser considerado para o próximo \texttt{MaxCommit}; índice do registo em votação para o próximo valor de \texttt{MaxCommit}.
\end{itemize}
Estas novas variáveis são atualizadas e partilhadas pelos processos nos pedidos \texttt{AppendEntries}, garantindo a coerência e convergência dos valores. Os processos usam o mapa de bits para descobrir quando uma maioria replicou corretamente o registo até \texttt{NextIndex}. Cada processo deve colocar o seu bit no \texttt{Bitmap} a ``um'' quando o seu registo possui a entrada em \texttt{NextIndex} e o mandato da última entrada é igual ao mandato atual.
Definimos duas funções para atualizar \texttt{NextCommit} e \texttt{MaxCommit}: \texttt{Update}, para atualizar os valores quando a maioria é alcançada no mapa de bits; e \texttt{Merge}, para combinar os valores recebidos de diferentes processos durante a propagação epidémica.
O algoritmo garante que \texttt{NextCommit} é maior que \texttt{MaxCommit} antes e após o uso das funções \texttt{Merge} e \texttt{Update}.

\begin{algorithm}[b!]
\SetAlgoLined
\SetKwData{bm}{bitmap}
\SetKwData{mc}{maxCommit}
\SetKwData{nc}{nextCommit}
\SetKwData{ci}{commitIndex}
\SetKwData{log}{log}
\SetKwFunction{max}{Max}
\SetKwFunction{min}{Min}
\SetKwFunction{update}{Update}
\SetArgSty{textrm}

\SetKwInput{KwData}{State}
\KwData{\bm; \mc; \nc; \ci}
\SetKwProg{Fn}{Function}{:}{}
\Fn{\update{}}{
    \everypar={\nl}
    \If{count of ``1"s in \bm higher or equal than majority size}{
        \mc $\leftarrow$ \nc\;
        \bm $\leftarrow$ \{0, 0, ..., 0\}\;
        \eIf{ \nc $\geq$ index of last entry \textbf{or} term of last entry is not equal to current term}{
            \nc $\leftarrow$ \nc + 1\;
        }{
            \nc $\leftarrow$ index of last entry\;
            \bm[$i$] $\leftarrow$ 1\;
        }
    }
}

\caption{Função para atualizar \texttt{Bitmap}, \texttt{NextCommit} e \texttt{MaxCommit} para o processo $P_i, i \in 0..n-1$}
\label{alg:update}
\end{algorithm}

A função \texttt{Update}, conforme o Algoritmo~\ref{alg:update}, atualiza os valores de \texttt{MaxCommit} e \texttt{NextCommit} e repõe o \texttt{Bitmap} a ``zeros'' quando este demonstra uma maioria (linha 1). Isto significa que uma maioria dos processos colocou ``um'' no \texttt{Bitmap}, ou seja, replicaram o seu registo até \texttt{NextCommit} e, então, \texttt{NextCommit} e \texttt{MaxCommit} podem avançar. \texttt{MaxCommit} recebe o valor de \texttt{NextCommit} e o \texttt{Bitmap} é reposto com ``zeros'' (linhas 2 e 3). 
Depois, \texttt{NextCommit} é atualizado conforme o estado atual do registo local: 
se \texttt{NextCommit} está num ponto mais avançado que o registo local ou o registo não possuí nenhuma entrada com o mandato atual (linha 4) então o valor de \texttt{NextCommit} é incrementado (linha 5); senão, o processo tem uma entrada a seguir ao \texttt{NextCommit} atual, com o mandato atual e, portanto, avança-se \texttt{NextCommit} para o índice da entrada mais avançada no registo (linha 7). Além disso, satisfazem-se as condições para colocar a ``um" o valor respetivo ao processo no \texttt{bitmap} (linha 8). 

A função \texttt{Merge} combina os valores com os recebidos de outros processos nos pedidos \texttt{AppendEntries} conforme o Algoritmo~\ref{alg:merge}. Primeiro, \texttt{MaxCommit} é atualizado para o maior entre o valor no estado e o recebido (linha 1). 
Depois, se \texttt{NextCommit} local for menor ou igual que o recebido significa que a informação no \texttt{Bitmap} recebido pode ser incluída no \texttt{Bitmap} local através de um ``ou lógico'' bit a bit (linhas 2 a 4). 
Caso uma maioria de processos tenha já replicado o registo até \texttt{NextCommit} (linha 5), é necessário substituir \texttt{Bitmap} e \texttt{NextCommit} com os recebidos (linhas 6 e 7), dado que estão mais avançados.  

\begin{algorithm}[t]
\SetAlgoLined
\SetKwData{bm}{bitmap}
\SetKwData{mc}{maxCommit}
\SetKwData{nc}{nextCommit}
\SetKwFunction{max}{Max}
\SetKwFunction{bor}{BitwiseOR}
\SetKwFunction{onecount}{OneCount}
\SetKwFunction{merge}{Merge}

\SetKwInput{KwData}{State}
\KwData{\bm; \mc; \nc}
\SetKwProg{Fn}{Function}{:}{}
\Fn{\merge{bitmap', maxCommit', nextCommit'}}{
    \everypar={\nl}
    \mc $\leftarrow$ \max{\mc, maxCommit'}\;
    \If{\nc $\leq$ nextCommit'}{
        \bm $\leftarrow$ \bor{\bm, bitmap'}\;
    }
    \If{\nc $\leq$ \mc}{
        \bm $\leftarrow$ bitmap'\;
        \nc $\leftarrow$ nextCommit'\;
    }
}
\caption{Função para combinar \texttt{bitmap}, \texttt{nextCommit} e \texttt{maxCommit} recebidos nos pedidos \texttt{AppendEntries} para o processo $P_i, i \in 0..n-1$}
\label{alg:merge}
\end{algorithm}

\begin{figure}[t]
    \centering
    \makebox[\textwidth]{
\includegraphics[width=1\textwidth]{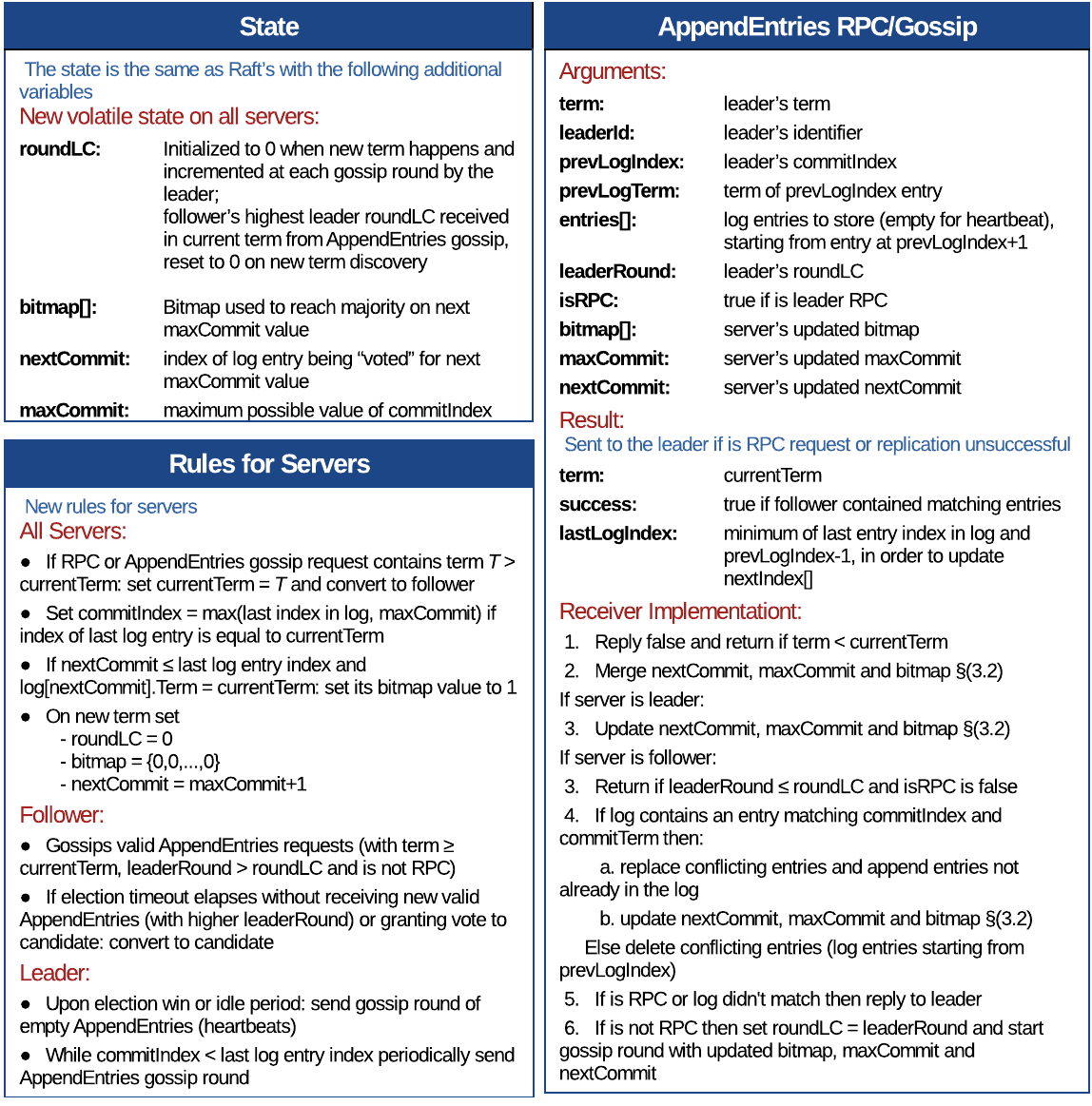}
    }
    \caption{Raft com propagação epidémica de \texttt{AppendEntries}}
    \label{fig:driftwood2}
\end{figure}

Os seguidores podem atualizar o seu \texttt{CommitIndex} para o mínimo entre o índice da última entrada no registo e o \texttt{MaxCommit} se o mandato da última entrada no registo for igual ao mandato atual.

Raft garante a coerência dos registos replicados, permitindo que o número de entradas nos registos difira transitoriamente entre processos. Ao atualizar \texttt{NextCommit} escolhe-se o maior valor possível conforme o estado do registo do processo. Contudo, um novo líder pode ser eleito com um registo menor que este \texttt{NextCommit}, o  que pode causar atrasos na confirmação de entradas. 
Para resolver este problema, as variáveis \texttt{NextCommit} e \texttt{Bitmap} devem ser redefinidas quando uma eleição é iniciada ou se se descobre que foi iniciado um novo mandato.
Processos, nestes casos, devem repor o \texttt{bitmap} a ``zeros'' e atribuir o valor de \texttt{MaxCommit+1} a \texttt{NextCommit}, pois é garantido pelo Raft que um processo só pode ser eleito líder se possuir o seu registo replicado até \texttt{MaxCommit}, uma vez que uma maioria de processos já tem o seu registo confirmado até esse ponto.

Caso o líder falhe sem ter descoberto que um seguidor encontrou um \texttt{MaxCommit} maior que o seu, o próximo processo eleito possuirá todas as entradas confirmadas, ou seja, as entradas do registo do líder até ao maior \texttt{MaxCommit} descoberto. Segundo o Raft, um processo pode ser eleito se o seu registo possuir todas as entradas confirmadas pelo líder. Um candidato é eleito se receber uma maioria de votos. Numa eleição, cada seguidor vota no primeiro candidato que pede o seu voto e possui o seu registo tão ou mais avançado que o seu, de acordo com a propriedade de correspondência de registo. O valor de \texttt{MaxCommit} indica que uma maioria de processos replicou com sucesso o registo do líder até este índice. Portanto pela restrição de eleição do Raft apenas candidatos que conseguiram replicar o seu registo até ao maior \texttt{MaxCommit} descoberto conseguirão receber uma maioria de votos e tornar-se líder. Assim qualquer novo possível líder possuirá todas as entradas confirmadas em termos anteriores.
A Figura~\ref{fig:driftwood2} sumariza as novas alterações do algoritmo relativamente ao Raft.

\section{Avaliação Experimental}
Fazemos uma análise comparativa entre as duas novas versões de Raft e o original. Para tal usamos Paxi \cite{paxi,paxiarticle}, uma infraestrutura para prototipagem e avaliação de algoritmos de replicação escrita em Go. 

\subsection{Configurações}
Corremos as experiências numa única máquina com 128 núcleos, incluindo os processos Raft (réplicas) e os clientes. Cada réplica corre em apenas um núcleo dedicado, para reduzir a variação dos resultados causado pela partilha de recursos. Usamos o cliente Paxi para produzir carga no sistema e realizar a avaliação. O cliente Paxi permite simular vários clientes concorrentes com ou sem uma taxa de pedidos determinada. Cada cliente envia um pedido e espera pela resposta, antes de enviar o próximo.
Os testes são executados apenas na fase de replicação do algoritmo com um líder estável.

Comparamos o desempenho, disponibilidade e escalabilidade do Raft original e das duas novas versões. Chamamos Versão 1 ao Raft com propagação epidémica de \texttt{AppendEntries} e Versão 2 à que usa também as novas estruturas de dados. Para analisar os algoritmos medimos: a latência média de resposta, 
taxa de processamento de pedidos, 
o uso de CPU por cada réplica 
e a latência entre as entradas recebidas pelo líder e a sua confirmação em cada réplica.
Corremos cada experiência 3 vezes e usamos a média dos valores resultantes no desenho dos diagramas.

\def\figscale{.49}
\def\figtrim{12 0 12 0}

\begin{figure}[b]
    \centering
\includegraphics[width=\figscale\textwidth,trim=24 0 24 0]{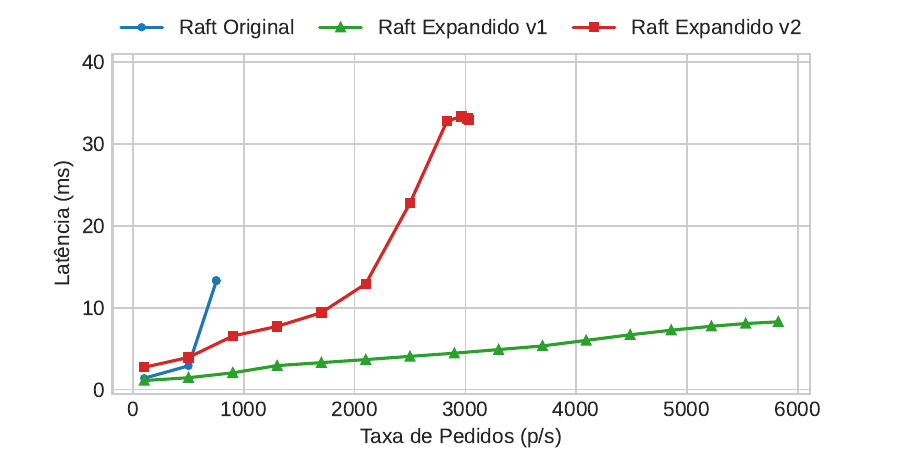}
    \caption{Latência média por taxa de pedidos.}
    \label{fig:lat_thr}
\end{figure}

\begin{figure}[t]
    \centering
    \begin{subfigure}[b]{\figscale\textwidth}
        \centering
        \includegraphics[width=\textwidth,trim=24 0 24 0]{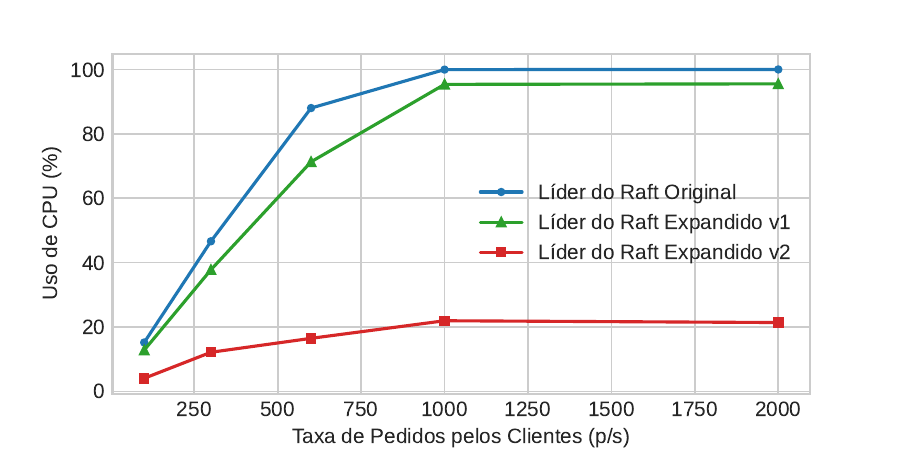}
        \caption{Uso do CPU pelos Líderes}
        \label{fig:cpurepleaders}
    \end{subfigure}
        \hfill
    \begin{subfigure}[b]{\figscale\textwidth}
        \centering
    \includegraphics[width=\textwidth,trim=24 0 24 0]{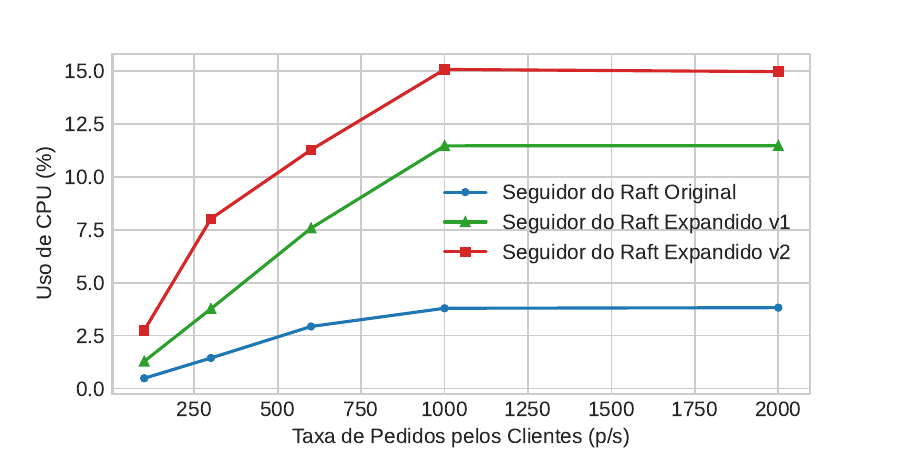}
        \caption{Uso do CPU pelos Seguidores}
        \label{fig:cpurepseguidores}
    \end{subfigure} 
    \caption{Uso de CPU por taxa de pedidos pelos clientes.}
    \label{fig:cputhr}
\end{figure}

\begin{figure}[t]
    \centering
    \begin{subfigure}[b]{\figscale\textwidth}
        \centering
        \includegraphics[width=\textwidth,trim=24 0 24 0]{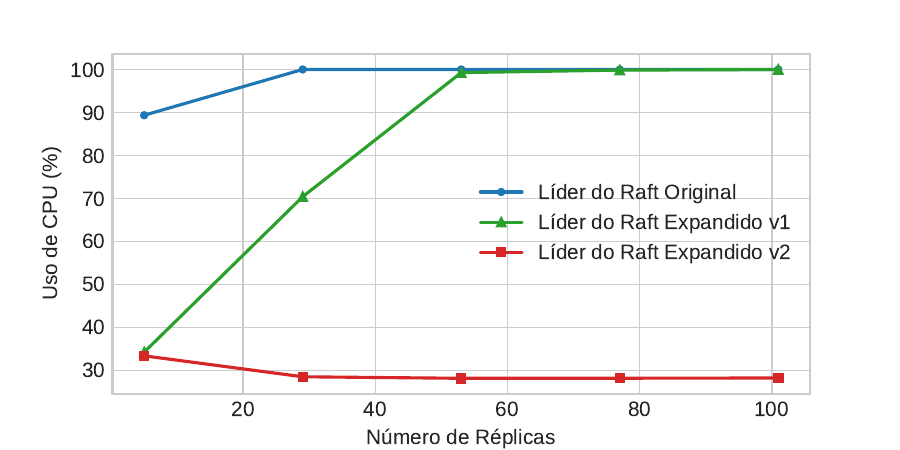}
        \caption{Uso do CPU pelos Líderes}
        \label{fig:cputhrleaders}
    \end{subfigure}
        \hfill
    \begin{subfigure}[b]{\figscale\textwidth}
        \centering
    \includegraphics[width=\textwidth,trim=24 0 24 0]{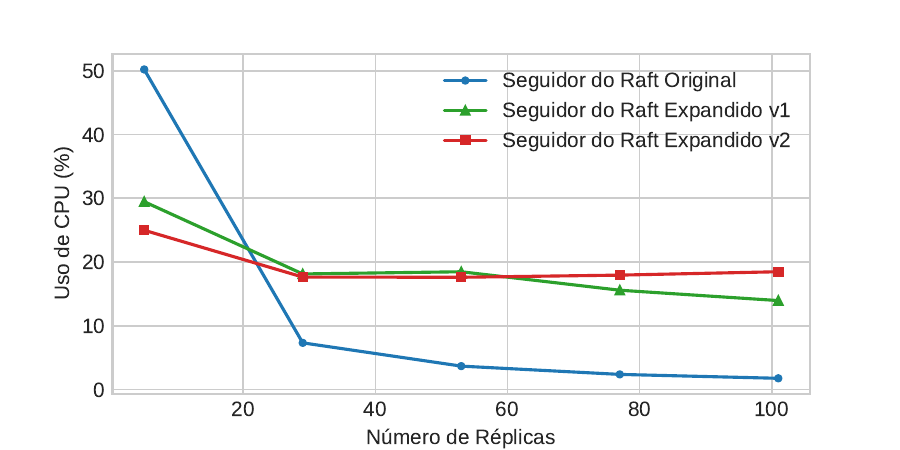}
        \caption{Uso do CPU pelos Seguidores}
        \label{fig:cputhrseguidores}
    \end{subfigure} 
    \caption{Uso de CPU por número de réplicas no sistema.}
    \label{fig:cpurep}
    \end{figure}

\subsection{Resultados}
Para analisar o desempenho corremos várias experiências com diferentes taxas de pedidos nos clientes. Em cada experiência usamos 100 clientes concorrentes e 51 réplicas.

A Fig.~\ref{fig:lat_thr} mostra os resultados da análise de desempenho dos algoritmos. Observamos que as novas versões apresentam um desempenho significativamente melhor do que o Raft original. A Versão 1 alcança uma taxa de pedidos muito maior com um pequeno aumento na latência. Contudo, a Versão 2 atinge o seu limite com uma certa taxa de pedidos, e a latência cresce num ritmo mais elevado, devido aos saltos necessários para fazer progredir as estruturas de dados trocadas e consequentemente o \texttt{CommitIndex}.

Analisamos também o uso de recursos, nomeadamente do CPU, pelos seguidores e pelo líder com diferentes cargas de trabalho e número de réplicas no sistema. O cliente do Paxi simula 10 clientes concorrentes que enviam pedidos imediatamente após receberem as respostas.

A Fig.~\ref{fig:cputhr} mostra o uso de CPU com o aumento da carga de trabalho pelo cliente do Paxi. Observamos que o líder da Versão 1 usou menos recursos que o líder do Raft original, causado pelo uso de propagação epidémica. A Versão 2 mostra um melhor uso dos recursos distribuídos, pois o líder não precisa esperar por respostas diretamente dos seguidores para avançar o \texttt{CommitIndex}. Comparativamente aos seguidores, o líder da Versão 2 tem ainda as tarefas adicionais de receber os pedidos dos clientes e iniciar a propagação epidémica, o que resulta ainda assim num uso da CPU ligeiramente superior aos seguidores.

Na Fig.~\ref{fig:cpurep} observamos que o Raft original é altamente centralizado no líder, pelo que uso de CPU pelo líder é muito maior que o dos seguidores. O uso de propagação epidémica resultou num melhor uso de recursos distribuídos. A Versão 1 escala melhor com o aumento de réplicas, mas tal como no Raft original o líder torna-se o gargalo do sistema ao chegar a um certo número de réplicas. Na Versão 2, o líder não se torna em nenhum ponto o gargalo do sistema, e o uso de CPU nas réplicas segue uma tendência quase linear, mostrando o melhor uso dos recursos distribuídos entre os três algoritmos.

\begin{figure}[t]
\centering
\begin{subfigure}[b]{\figscale\textwidth}
    \centering
    \includegraphics[width=\textwidth,trim=24 0 24 0]{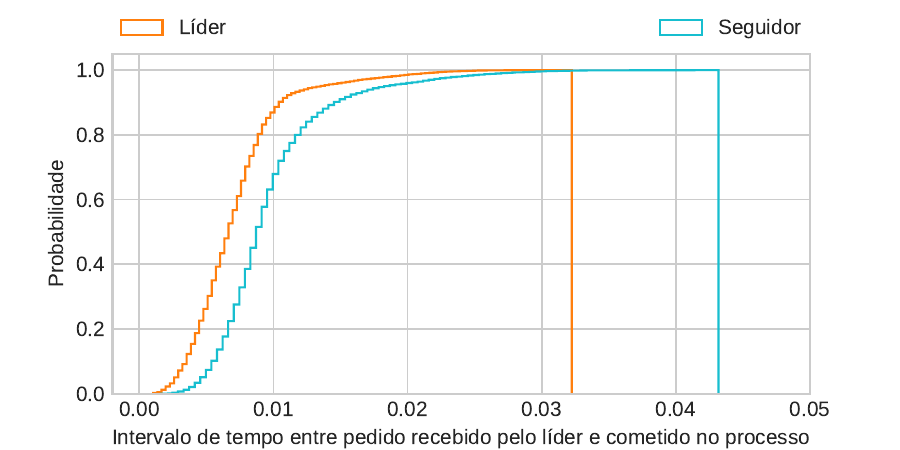}
    \caption{Original}
    \label{fig:raftcdf}
\end{subfigure}
\hfill
\begin{subfigure}[b]{\figscale\textwidth}
    \centering
\includegraphics[width=\textwidth,trim=24 0 24 0]{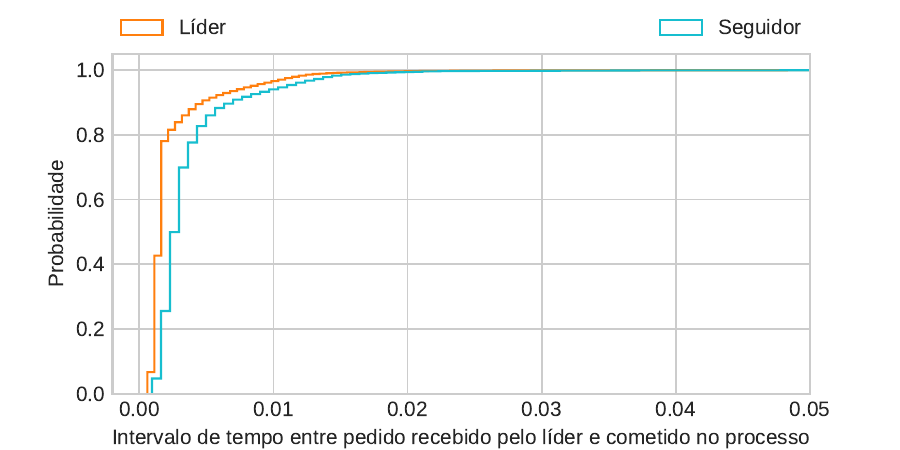}
    \caption{Versão 1}
    \label{fig:v1cdf}
\end{subfigure} 
    \hfill
\begin{subfigure}[b]{\figscale\textwidth}
    \centering
\includegraphics[width=\textwidth,trim=24 0 24 0]{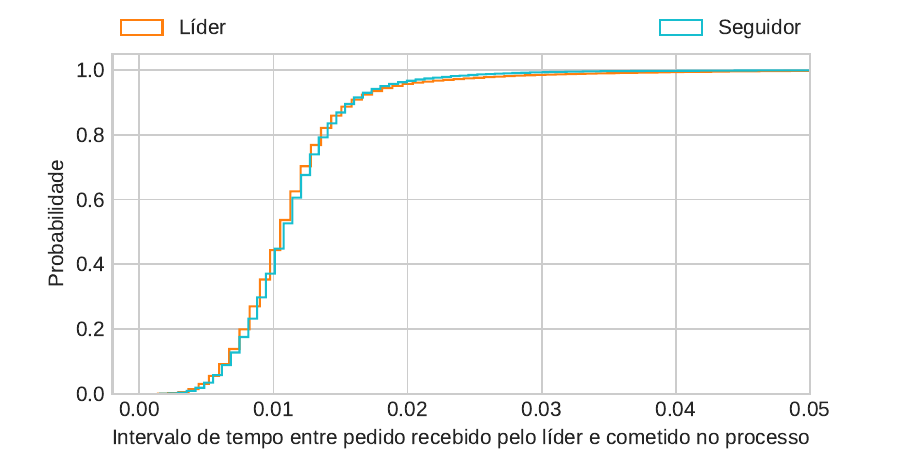}
    \caption{Versão 2}
    \label{fig:v2cdf}
\end{subfigure} 
\caption{FDA do intervalo de tempo entre pedido do cliente recebido pelo líder e confirmado na réplica.}
\label{fig:cdf}
\end{figure}

Finalmente, medimos a latência entre o momento em que um pedido é confirmado pelo líder e o momento em que é confirmado pelos seguidores. Uma baixa latência aqui é útil pois, caso os clientes estejam dispostos a relaxar as propriedades de coerência, é possível fazer leituras mais rápidas diretamente das réplicas.
A Fig.~\ref{fig:cdf} mostra a função de distribuição acumulada (FDA) do tempo que as réplicas demoram a confirmar os pedidos após a receção destes pelo líder. Observamos que os seguidores do Raft original e da Versão 1 demoram mais a confirmar os pedidos, pois o líder é responsável por informar os seguidores do seu \texttt{CommitIndex}. Por outro lado, a Versão 2 permite aos seguidores progredirem o \texttt{CommitIndex} sem necessitar da confirmação do líder, permitindo que o \texttt{CommitIndex} dum seguidor possa estar à frente do líder. Verificamos assim que a Versão 1 consegue ser mais rápida que a original, dada a menor carga. No caso da Versão 2, a decisão descentralizada permite que as réplicas não tenham latência adicional.

\section{Trabalho relacionado}

Sendo um problema tão importante para sistemas distribuídos tolerantes a faltas, existem inúmeras propostas de algoritmos de consenso, incluindo até alguns que fazem também utilização de propagação epidémica. Concretamente, a primeira proposta nesse sentido \cite{mutable,mmpo11}, que introduz a utilização de uma permutação dos destinos de cada mensagem para conciliar a aleatoriedade com o determinismo, baseia-se no algoritmo de Chandra e Toueg \cite{10.1145/226643.226647} (que se destina a tomar apenas uma decisão e não a ordenar uma sequência) pelo que não é aplicável ao registo do Raft. 

Mais recentemente, a propagação epidémica foi utilizada no algoritmo Semantic Gossip \cite{Cason2021-ka}, de forma semelhante ao Mutable Consensus \cite{mutable} mas aplicada ao Paxos. Neste caso, não é usada uma permutação dos destinos, pelo que o algoritmo se torna probabilista. 

Uma estratégia alternativa para evitar o gargalo na difusão das mensagens pelo líder consiste na pré-difusão dos comandos \cite{6424845,Tennage2022-dq} e depois na decisão da sua ordem. Esta abordagem torna no entanto o protocolo mais complicado, pois obriga a relacionar as garantias da pré-difusão com as do consenso e não melhora o processo de recolha de votos.

Os algoritmos de consenso podem também ser descentralizados de forma a eliminar completamente o líder. Neste caso, é acrescentada meta-informação aos comandos sobre os conflitos, de forma a que comandos que não interfiram possam ser ordenados concorrentemente por diferentes processos produzindo assim uma ordem parcial \cite{Moraru2013-do}. No entanto, cada processo tem ainda que comunicar diretamente com todos os outros.

Finalmente, a propagação epidémica é também usada nos protocolos de acordo bizantino, normalmente no contexto de \emph{blockchains}, para a propagação das transações \cite{tendermint}. Neste caso, a possível utilização de propagação epidémica para recolha de votos é substancialmente mais complicada \cite{sapo20}.

\section{Conclusões}
A propagação epidémica permitiu descentralizar a replicação do registo do líder, evitando que este se torne o gargalo do sistema, e distribuir de forma mais equilibrada a carga de trabalho entre os processos, resultando num melhor desempenho e uso dos recursos distribuídos. Nomeadamente, a Versão 1 do algoritmo apresentado permitiu aumentar $6\times$ o débito máximo atingível e a Versão 2 diminuir para $1/3$ a carga de CPU do líder, ambos em cenários com 51 réplicas. 

As novas versões do Raft deverão também ser resilientes em cenários complexos de rede.
Como trabalho futuro, será importante avaliar experimentalmente os algoritmos em cenários do mundo real através da \textit{cloud}, com experiências em redes locais (LAN) e redes de larga escala (WAN). Além disso, pretendemos expandir mais os algoritmos para os tornar mais robustos em redes não transitivas. Por exemplo, usar propagação epidémica para recolher votos durante a eleição ou criar mecanismos que permitam seguidores mais atualizados replicar as suas entradas noutros seguidores atrasados que podem ter dificuldades a comunicar com o líder.

\bibliographystyle{splncs04}
\bibliography{main}

\end{document}